# Influence of Bubble Lifetime on the Drying of Catalytically Active Sessile Droplets

Meneka Banik and Ranjini Bandyopadhyay*

Soft Condensed Matter Group, Raman Research Institute, C. V. Raman Avenue, Sadashivanagar, Bangalore 560080, India.

Corresponding author: *ranjini@rri.res.in

**ABSTRACT**

When colloidal droplets evaporate, suspended particles are redistributed by a competition between evaporation-driven capillary advection, interfacial Marangoni stresses and particle mobility, leading to diverse deposition patterns relevant to coating and self-assembly. While these mechanisms are well understood for passive suspensions, their interplay in chemically active colloidal systems remains less explored. Here, we investigate the drying dynamics of aqueous droplets containing catalytic polystyrene-platinum (PS-Pt) Janus particles in the presence of hydrogen peroxide ($H_2O_2$) fuel. $H_2O_2$ undergoes catalytic decomposition at the Pt hemisphere, resulting in the generation of oxygen ($O_2$). By systematically varying $H_2O_2$ concentration, surface wettability and open versus confined drying conditions, we identify distinct transport regimes governed by the interplay between capillary flow and an inward flow driven by activity-induced surface tension gradients. While time-resolved contact angle measurements reveal substrate-dependent evaporation modes, increase in catalytic activity promotes $O_2$ bubble generation that locally reverses or disrupts outward particle transport. Closed drying conditions further modify evaporation rates and prolong bubble lifetimes, leading to transitions from peripheral accumulation to spatially uniform or centrally concentrated deposits. Bubble lifetime, controlled here by tuning substrate wettability and $H_2O_2$ concentration, therefore emerges as the dominant mechanism governing the dried morphologies of catalytically active Janus particle droplets.

**KEYWORDS:**

active particles; capillary advection; colloidal droplets; evaporative self-assembly; Marangoni flow; wettability

## INTRODUCTION

The drying of colloidal droplets on flat substrates has long served as a canonical model system for probing interfacial phenomena such as capillarity, wetting transitions and particle transport at confined liquid-vapour interfaces [1-4]. In the classical framework, passive colloidal suspension droplets evaporating under isothermal conditions form ring-like or heterogeneous deposits governed primarily by an outward capillary flux directed towards pinned contact lines [1,2]. Subsequent studies demonstrated that thermal or solutal Marangoni stresses may partially counteract this radial flow, generating internal recirculation that redistributes liquid and suspended particles [5-7]. Together, these principles underpin strategies for producing reproducible coatings, inkjet-printed films, and functional thin layers [8-10]. However, these models rely on the central assumption that suspended particles behave as passive tracers that do not themselves modify the evolving interface or the evaporation pathway [11,12]. This assumption becomes questionable when particles possess intrinsic surface activity or anisotropic interfacial affinity [13].

Janus colloids introduce a fundamentally different scenario, as their chemically asymmetric surfaces interact differently with the liquid-vapour and solid-liquid interfaces [14,15]. For example, in polymer-metal Janus particles, the hydrophilic metal cap and hydrophobic polymer body produce heterogeneous adhesion, orientation, and mobility that depend strongly on substrate wettability [16-18]. For catalytically active Janus particle droplets, substrate wettability also governs whether evaporation-driven fuel enrichment or reaction-driven depletion dominates near the contact line [19-21]. Unlike passive systems, active droplets exhibit localised catalytic consumption at the three-phase boundary which induce inward Marangoni flows and redistribute particles [22]. The competition between these mechanisms determines particle segregation, clustering, and overall deposit uniformity even in the absence of sustained chemical fluxes, highlighting the critical role of particle-substrate interactions [20,23,24]. Similar coupling between activity and evaporation-driven transport has been observed in bacterial systems, where motility modifies contact line dynamics and deposition patterns in drying droplets [25].

In active systems, self-generated chemical fluxes continuously perturb the interfacial environment [26,27]. First, substrate surface energy controls contact line mobility and pinning strength, setting the length scale for interfacial Marangoni stresses [5,28]. Strong pinning promotes stress accumulation, whereas weak pinning allows contact line motion and partial stress relaxation [29,30]. Second, hydrophilic substrates promote droplet spreading into thin precursor films with shallow curvature and enhanced liquid-air interfacial area [31,32], while hydrophobic substrates preserve curved droplet footprints and their prolonged interfacial residence times [33,34].

In polystyrene-platinum (PS-Pt) Janus particle aqueous droplets containing hydrogen peroxide ($H_2O_2$) fuel, catalytic decomposition at the Pt cap generates dissolved oxygen ($O_2$) that nucleates into bubbles at higher fuel concentrations. The growth, residence, and detachment of these bubbles depend on wettability, confinement, and particle interactions [6,35,36]. Hydrophobic substrates enhance interfacial mobility and intermittency, whereas hydrophilic substrates promote stronger pinning and gradual stress relaxation [37], making bubble-mediated perturbations a consequence of wettability-controlled interfacial dynamics [38-40]. In this work, we identify bubble lifetime as the key parameter governing evaporation pathways in active Janus droplets. By systematically varying substrate wettability, fuel concentration, and evaporation environment, we control bubble lifetime and quantify its impact on contact line dynamics, interfacial stress relaxation, and particle transport. Time-resolved contact angle measurements and final deposit morphologies reveal how Janus polarity couples with wettability to select distinct evaporation modes. These results establish a unified framework linking substrate chemistry, bubble dynamics, and active assembly, providing design guidelines for coating, printing, and microreactor applications.

## MATERIALS AND METHODS

***Janus particle synthesis***: Polystyrene (PS) microspheres with a mean diameter of 2.07 ± 0.15 μm (1% solids, Fluorescent PS, Dragon Green, Bangs Laboratories Inc.) were used for Janus particle fabrication. The suspension of PS microspheres was diluted in isopropyl alcohol (IPA, Sigma-Aldrich, ≥ 99.7%, FCC, FG) in the ratio of 1:4. Clean glass substrates (Blue Star, thickness 1.35 mm) were prepared by sequential ultrasonication in acetone (Sigma-Aldrich, ACS reagent, ≥ 99.5%), ethanol (RCP Distilleries India Pvt. Ltd., 99.9%, Analytical Grade), and Milli-Q water (resistivity: 18.2 MΩ.cm at 25 °C, < 5ppb TDS) (15 min each), followed by treatment in piranha solution (3:1 mixture of $H_2SO_4$, Merck, 98%, for analysis) and $H_2O_2$ (Thermo Fisher Scientific, 30 wt%)). The substrates were rinsed thoroughly with Milli-Q water, dried in a hot air oven, and purged with nitrogen before use. 200 μL of the diluted PS suspension was dispensed onto a cleaned glass slide and kept undisturbed in a clean Petri dish at 25 °C for 8 hours [41]. A close-packed monolayer of PS microspheres, spanning an area of approximately 100 μm$^2$, was obtained on the cleaned glass slides by sedimentation (Supplementary Material Figure S1a). A 20 nm thick platinum (Pt) film was then deposited onto the exposed hemispheres of the PS particles using a benchtop sputter coater (Quorum Q150RS), which resulted in half-coated PS particles (Supplementary Material Figure S1b). The Janus particles were released from the substrate into Milli-Q water via ultrasonication for 1 minute (Maxsell ultrasonic cleaner, MX150QTD-6L, power 150 W), yielding a stock suspension. The stock suspension was washed with Milli-Q water three times to remove any Pt debris by repeated centrifugation (using Tarsons SPINWINTM MC 03 microcentrifuge at 5000 rpm for 5 minutes at room temperature), followed by re-dispersion. This washed suspension was once again centrifuged, the supernatant discarded, and the particles redispersed in Milli-Q water to obtain a 1 wt% aqueous suspension of Janus particles [13,42]. The Pt coating on the PS particles was confirmed by dropcasting 1 μl of the dispersion on clean glass slide (Supplementary Material Figure S1c). To induce catalytic activity on the Pt side, appropriate quantities of $H_2O_2$ were added to the final Janus suspension, yielding overall $H_2O_2$ concentrations between 0.1 and 10 wt% in the suspension medium.

***Substrate preparation***: Clean silicon wafers were used as the hydrophilic substrate. For cleaning, the substrates were immersed in piranha solution (3:1 mixture of $H_2SO_4$ and $H_2O_2$), followed by thorough rinsing with Milli-Q water and drying with nitrogen gas before use. Hydrophobic substrates were fabricated using Sylgard 184 (a two-part cross-linkable polydimethylsiloxane (PDMS) elastomer; Dow Corning, USA) films. The oligomer (part A) to cross-linker (part B) ratio was maintained at 10:1 (wt/wt). The Sylgard mixture was poured into a clean glass Petri dish and cured at 120 °C for 12 hours in a hot air oven for complete crosslinking, after which it was carefully peeled off [43] and cut into desired sizes. Atomic force microscopy (AFM) was used to quantify the surface roughness of the hydrophilic and hydrophobic (PDMS) substrates. Both surfaces exhibited low nanoscale roughness (~2 nm), confirming minimal topographical heterogeneity and reducing the likelihood of contact line pinning arising from surface defects (Supplementary Material Figure S2a and b).

***Droplet drying:*** All drying experiments were conducted under controlled ambient conditions of 23 °C and 40 % relative humidity. We employed two distinct drying conditions for our experiments: open (where the substrate was kept open to the environment) and closed (where the substrate was placed in a closed Petri dish). 2 μL suspension droplets were carefully dispensed onto the substrates using a calibrated micropipette (Tarson). Closed drying experiments were performed by placing the substrate with the dispensed droplet inside a Petri dish (diameter: 10 cm), which was then covered to restrict vapour exchange with the surroundings. The enclosure was not actively humidity-controlled; however, confinement of evaporated vapour led to a progressive increase in local relative humidity, thereby reducing the evaporation rate compared to open conditions. All closed environment experiments were conducted using the same Petri dish geometry and setup to ensure consistency and reproducibility across different $H_2O_2$ concentrations and substrate types. The drying dynamics were monitored in situ using a stereo zoom microscope (Olympus SZX16). Simultaneously, evolution of the droplet contact angle was recorded using a contact angle goniometer (Attension Theta Flex- Biolin Scientific Optical Tensiometer) in sessile droplet mode, enabling quantitative analysis of wetting behaviour and evaporation kinetics. Droplet images were acquired at a frame rate of 2.3 frames per second, and were analysed using the OneAttension software. After complete evaporation, the final dried morphologies were imaged using scanning electron microscopy (SEM, Ultra Plus 4098 FESEM, Carl Zeiss) and atomic force microscopy (AFM, Oxford Instruments MFP-3D-bio-AFM in tapping mode, with resonant frequency of 157 kHz). All experiments were repeated a minimum of five times to ensure reproducibility of the results. A schematic overview of the experimental system and measured quantities is provided in Figure 1. The key experimental parameters systematically varied include substrate wettability, $H_2O_2$ concentration, and evaporation environment. The evaporation process was quantitatively characterised through contact angle measurements, while the final deposition patterns were analysed in terms of morphology, circularity, and ring thickness.

***Analyses:*** Quantitative analyses of the evaporation dynamics and final deposit morphologies were performed using image-based methods. Circularity was quantified in MATLAB by tracing the outer boundary of the dried deposit and computing the standard shape factor $C = 4\pi A/P^2$ (computed using the function `regionprops`), where $A$ is the projected area and $P$ is the perimeter (with $C$=1 corresponding to a perfect circle). Ring thickness was also determined using a MATLAB script by defining the outer and inner annular boundaries and calculating their radial separation, averaged over 20 measurements to account for local heterogeneity. Bubble lifetime was defined as the duration between the first appearance and complete disappearance (detachment or dissolution) of an individual bubble, measured from time-resolved side-view contact angle videos. Bubble count was obtained by tracking the number of bubbles for up to 5 minutes after dispensing the droplet. This was quantified from top-view optical microscope videos, at time intervals of 30 seconds. All measurements were averaged over multiple independent experiments to ensure reproducibility. These measurements were done for an open system, since the closed Petri dish caused imaging difficulties.

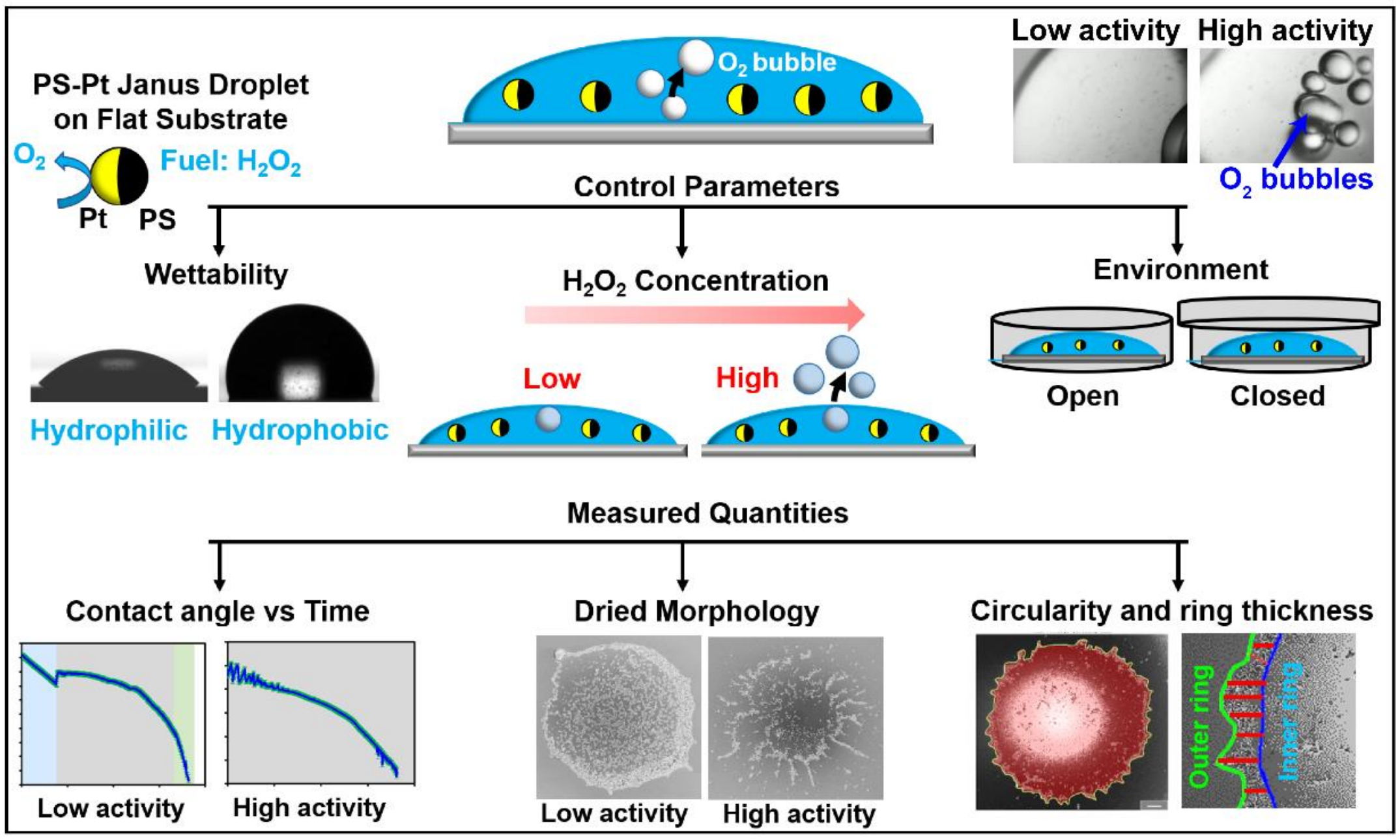

**Figure 1** Schematic illustration of evaporation dynamics in catalytically active PS-Pt Janus droplets on flat substrates. The system is controlled by substrate wettability (hydrophilic vs hydrophobic), fuel concentration, and evaporation environment (open vs closed), which together regulate bubble dynamics and interfacial stress evolution. The resulting effects are quantified through time-resolved contact angle measurements and final deposition morphologies captured using scanning electron microscopy. The deposit morphologies are characterised by metrics such as circularity and ring thickness.

## RESULTS AND DISCUSSION

### *Evaporation dynamics of passive Janus particle droplets on hydrophilic and hydrophobic flat substrates:*

Figure 2 compares the evaporation dynamics and final deposit morphologies of droplets containing PS-Pt Janus particles in the absence of $H_2O_2$ fuel on hydrophilic and hydrophobic flat substrates. Under these conditions, particle motion arises solely from evaporation-driven capillary flows. As presented in Figures 2a1,a2,b1,b2, the entire drying regime can be divided into four zones: constant contact radius (CCR) mode (indicated in blue), constant contact angle (CCA) mode (indicated in yellow), mixed mode (indicated in grey) and fast evaporation step (indicated in green) for both hydrophilic and hydrophobic flat substrates.

On hydrophilic substrates (Figures 2a1-a2), droplets initially evaporate in a pinned CCR mode, as indicated by a gradual decrease in contact angle accompanied by minimal variation in diameter of the droplet footprint (base diameter of the droplet on the surface immediately after dispensing). The strong solid-liquid affinity promotes rapid spreading and stabilises the three-phase contact line, producing a shallow droplet profile. As evaporation progresses, depinning occurs eventually and the droplet transitions to a constant contact angle (CCA) regime. Although the substrate is hydrophilic, depinning is facilitated by the progressive reorientation of the Janus particles near the contact line, as described in previous works [14,18]. Initially, particles anchor (pin) with their polymeric PS hemispheres on the substrate, stabilising the three-phase contact line. As the droplet thins, the wettability contrast between PS and Pt generates asymmetric surface tension forces that drive reorientation, allowing the Pt caps to contact the substrate. Once sufficient particles reorient, contact line depinning occurs despite the surface being hydrophilic. This evaporation dynamics is directly reflected in the final dried morphologies observed here (Figure 2a3-a4). Inspection of the acquired SEM image (Figure 2a4) after droplet drying reveals preferential exposure of the polymeric PS hemisphere, indicating that the Pt side contacts the substrate (the brighter and rough hemisphere is identified as the Pt side, and the darker smooth hemisphere as the PS side).

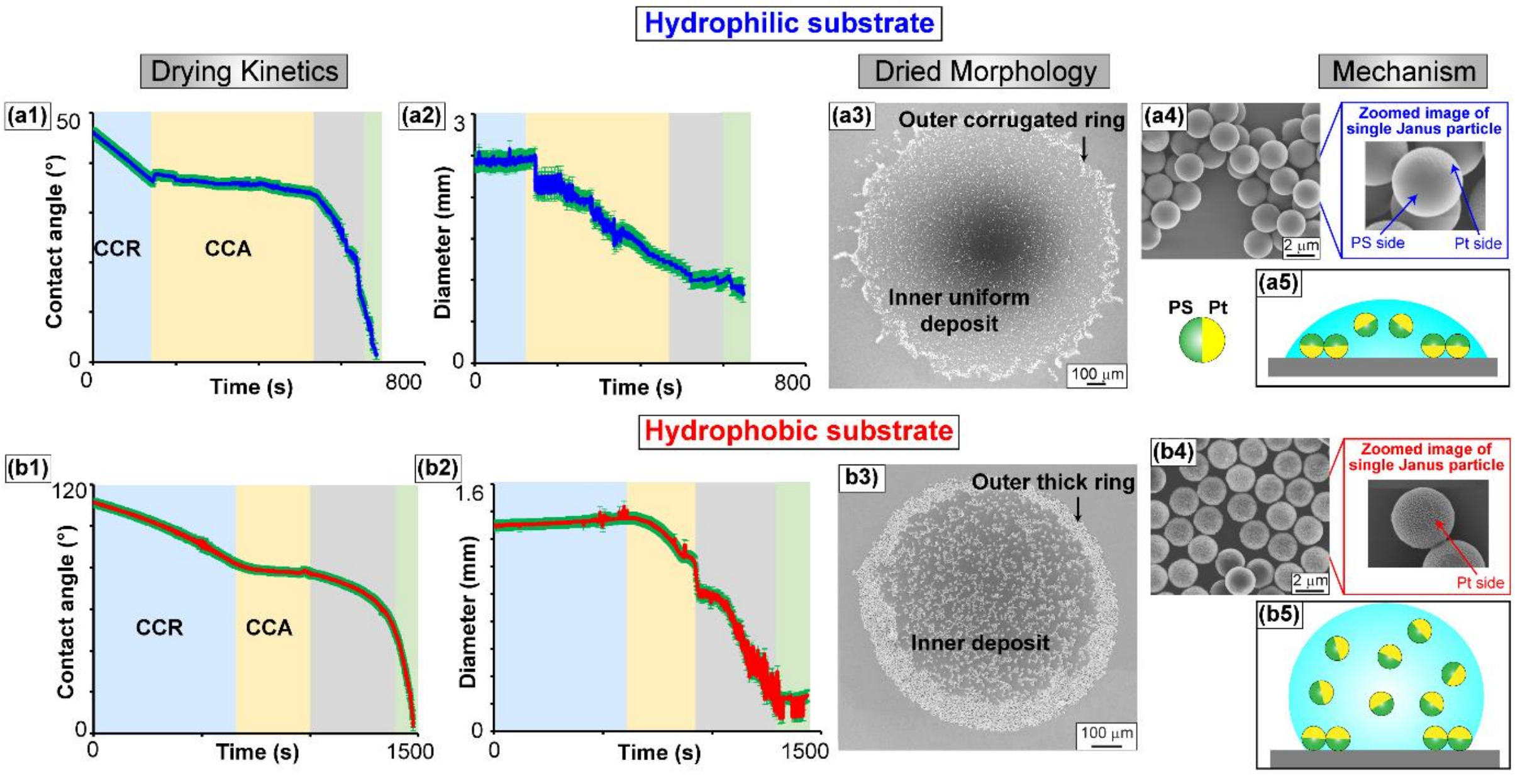

**Figure 2** Evaporation dynamics and final dried morphologies of evaporating PS-Pt Janus particle droplets on (a) hydrophilic and (b) hydrophobic substrates in the absence of $H_2O_2$ fuel. (a1,b1) Evolution of droplet contact angle with time, highlighting pinned (CCR) and depinned (CCA) stages. The error bars are highlighted in green. (a2,b2) Corresponding changes in droplet diameter with time. (a3,b3) SEM images of the final dried deposits on hydrophilic and hydrophobic substrates. (a4,b4) Magnified SEM images showing orientations of the Janus particles, with zoomed in images of single Janus particles highlighting the PS and Pt sides, (a5,b5) schematic representation of particle orientation on hydrophilic and hydrophobic substrates.

In contrast, droplets evaporating on hydrophobic substrates exhibit markedly different behaviour (Figures 2b1-b2). The initial contact angle is substantially larger, producing a taller and more curved droplet profile in which particles near the base are positioned closer to the triple phase contact line from the outset [16]. This enables rapid reorientation of the Janus particles such that the lower-energy polymeric PS hemispheres preferentially contact the substrate while the Pt caps remain exposed toward the air-liquid interface [44]. This early orientation allows the droplet to retain a pinned footprint for a significantly longer fraction of the drying process (blue region in Figures 2b1-b2). Depinning eventually occurs abruptly (yellow regions in the same figures). The prolonged pinned regime enhances the outward capillary flux, resulting in a thicker, more continuous outer ring in the final deposit (Figure 2b3). Corresponding SEM images reveal predominant exposure of the Pt hemispheres (Figure 2b4). The emergence of a mixed evaporation regime (grey region) in both cases arises from intermittent stick-slip motion driven by the orientation-dependent rearrangement of Janus particles at the contact line, causing both the contact angle and droplet footprint to decrease simultaneously before the final rapid evaporation stage. Notably, even in the absence of catalytic activity, the intrinsic anisotropy of Janus particles introduces an additional control pathway in droplet drying by enabling orientation-dependent adhesion and localised stress relaxation at the contact line.

***Fuel-driven evaporation dynamics of catalytically active Janus particle droplets on hydrophilic substrates under open drying conditions*:**

Figure 3 displays modifications in the evaporation dynamics of active PS-Pt Janus particle droplets on a hydrophilic flat substrate due to an increase in $H_2O_2$ concentration (0.1-10 wt % in Milli-Q water) under open drying conditions. The drying kinetics evolve systematically with $H_2O_2$ concentration (Figures 3a-f). The entire drying regime has been divided into three zones: CCR mode (indicated in blue), mixed mode (indicated in grey) and fast evaporation (indicated in green). The magnified side-views of the evaporating droplets (insets), captured with a contact angle goniometer, show the emergence of $O_2$ bubbles (highlighted in the inset of Figure 3b). Changes in the drying kinetics are mirrored in the intermediate top-view droplet videos captured with an optical microscope (snapshots presented in Supplementary Material Figure S3).

At low fuel concentrations (0.1-1 wt%, Figures 3a-b), the initial stage of evaporation is governed by capillary transport, during which time the droplet remains pinned and follows a CCR mode. This capillary flux sustains particle advection and produces an initially smooth decrease in contact angle. However, unlike the purely passive droplet in Figure 3a, where the interface transitions from CCR to a CCA regime, the active system does not exhibit a clear

CCA stage. Instead, the evaporation pathway enters a mixed mode following the initial pinned interval. This deviation arises because catalytic $O_2$ generation at the platinum caps begins to intermittently perturb the interface. Short-lived bubbles nucleate near the air-liquid interface (Figure 3b). Their repeated formation disrupts the uniform stress distribution required to sustain a stable CCA regime. Bubble nucleation induces non-uniform evaporation and local compositional gradients around the bubble, generating surface-tension differences that drive Marangoni flows. Owing to droplet curvature and buoyancy, bubbles preferentially localise near the apex, producing a flow that transiently opposes the outward capillary flux and advects particles toward the droplet centre [19]. Consequently, the contact angle data displays minor oscillations instead of a smooth trajectory. Capillarity still dominates the global drying dynamics in this low activity regime, but the interface is no longer purely passive. The combined action of outward evaporation-driven flow and intermittent bubble-mediated Marangoni recirculation produces a hybrid evaporation pathway.

As the $H_2O_2$ concentration increases to intermediate levels (3-5 wt%, Figures 3c-d), measurable deviations from classical evaporation become apparent. The contact angle curves develop fluctuations superimposed on the monotonic decay, indicating intermittent coupling between catalytic activity and interfacial curvature. Bubble nucleation becomes more frequent, and short-lived interfacial disturbances periodically modulate local stress distributions. This regime represents a transitional state in which catalytic $O_2$ generation begins to compete with evaporation-induced capillary stresses. The contact line still exhibits overall stability, but local pinning-depinning-repinning events become increasingly evident. At higher fuel concentrations (7.5-10 wt%, Figures 3e-f), the contact angle evolution displays pronounced, rapid, intermittent fluctuations during both early and intermediate drying stages. These oscillations correspond to repeated cycles of bubble nucleation, growth, coalescence and bursts at or near the air-liquid interface. The magnitude of catalytic activity becomes sufficient to repeatedly perturb the droplet interface, producing a highly irregular evaporation trajectory characterised by rapid curvature fluctuations and intermittent local depinning events. In this regime, bubble dynamics and associated interfacial stresses induce transient distortions of the liquid-air interface and modify the local force balance near the contact line. These perturbations can facilitate depinning events by overcoming local pinning barriers. Unlike the passive hydrophilic case, where evaporation proceeds through well-defined CCR and CCA regimes governed primarily by capillary flow and pinning forces, the Janus system exhibits activity-dependent deviations from classical behaviour. Even at low fuel concentrations, bubble-mediated interfacial perturbations suppress the emergence of a stable CCA stage, while increasing catalytic activity enhances contact angle fluctuations and promotes intermittent depinning. Thus, catalytic $O_2$ generation introduces a dynamic, non-equilibrium perturbation that competes with evaporation-driven capillarity and modifies the drying pathway.

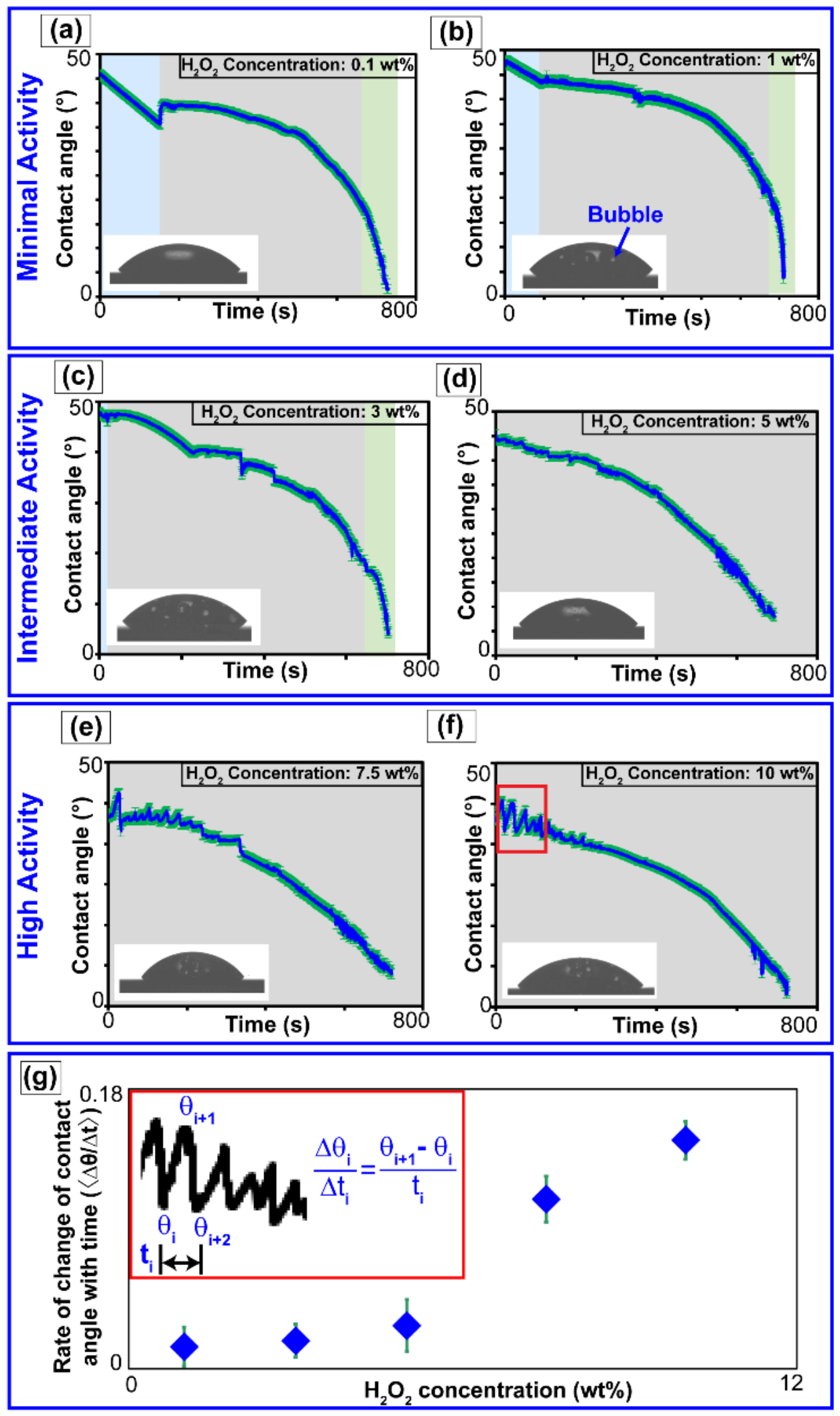

**Figure 3** Drying kinetics of catalytically active Janus particle droplets on a flat hydrophilic substrate with increasing $H_2O_2$ concentration: (a) 0.1 wt%, (b) 1 wt%, (c) 3 wt%, (d) 5 wt%, (e) 7.5 wt%, and (f) 10 wt%. (a)-(f): Evolution of the contact angle ($\theta$) with time. The error bars calculated from five individual runs are highlighted in green. Insets show side views of the evaporating droplets. (g) The rate of change of contact angle with time $\langle\frac{\Delta\theta}{\Delta t}\rangle$. Inset shows a schematic of the protocol followed to estimate $\langle\frac{\Delta\theta}{\Delta t}\rangle$.

To quantify the degree of interfacial disturbance, the rate of change of contact angle with time $\langle\frac{\Delta\theta}{\Delta t}\rangle$ was extracted from the contact angle trajectories and plotted as a function of $H_2O_2$ concentration (Figure 3g). $\langle\frac{\Delta\theta}{\Delta t}\rangle$ was chosen because it directly captures the temporal response of the droplet interface to force imbalances. In a capillary-dominated evaporation regime, the interface evolves smoothly. In contrast, when additional stresses are introduced, such as those arising from bubble growth, local pressure buildup, and surface tension gradients, the interface undergoes rapid, discrete curvature changes. Quantitatively, the contact angle is a measure of interface curvature at the three-phase boundary, which is governed by a balance of stresses. Any sudden change in these stresses due to bubble growth or detachment induces a rapid change in curvature, and hence a measurable change in contact angle. While $\langle\frac{\Delta\theta}{\Delta t}\rangle$ is not a direct measurement of Marangoni stress or velocity fields, it provides a robust, experimentally accessible measure of interfacial dynamics. As illustrated schematically in the inset of Figure 3g, each oscillatory rise and drop in the contact angle ($\theta$) was treated as an individual event. The magnitude of change between two successive extrema ($\theta_{i+1} - \theta_i$) was divided by the corresponding time interval ($t_i$), yielding a local $\Delta\theta_i/\Delta t_i$ for the *i*th segment. This procedure was repeated across multiple oscillation cycles, generating a set of discrete $\Delta\theta_i/\Delta t_i$ values. The final reported rate for a given $H_2O_2$ fuel concentration was obtained by averaging these segment-wise $\Delta\theta_i/\Delta t_i$ values over all identified oscillation segments for a given condition: $\langle\frac{\Delta\theta}{\Delta t}\rangle = \frac{1}{N}\sum_{i=1}^{N}\frac{\theta_{i+1}-\theta_i}{t_i}$

The resulting trend obtained from our experiments reveals a systematic increase in the average $\langle\frac{\Delta\theta}{\Delta t}\rangle$ with fuel concentration, indicating that interfacial curvature fluctuations become faster as catalytic $O_2$ generation intensifies. Overall, Figure 3 reveals a hierarchy of competing time scales. At low $H_2O_2$ concentrations, evaporation dominates and the droplet follows a primarily capillary-controlled pathway, whereas at intermediate concentrations, catalytic generation produces intermittent oscillations. At high concentrations, bubble generation overwhelms both processes, leading to sustained interfacial instabilities and a clear transition from capillary-dominated to catalytically-perturbed evaporation.

***Fuel-dependent evaporation dynamics of catalytically active Janus droplets on hydrophobic substrates under open drying conditions:***

Figure 4 quantifies how increasing $H_2O_2$ (fuel) concentration (0.1-10 wt% in Milli-Q water) progressively modifies the evaporation dynamics of PS-Pt active Janus particle droplets on a hydrophobic flat substrate, under open drying conditions. The drying kinetics evolve systematically with $H_2O_2$ concentration (Figures 4a-f). The entire drying regime has been divided into three zones: CCR mode (indicated in blue), mixed mode (indicated in grey) and fast evaporation (indicated in green). Insets show the magnified side-views of the evaporating droplets captured with a contact angle goniometer. Changes in the drying kinetics are mirrored in the intermediate top-view droplet videos captured with an optical microscope (snapshots in Supplementary Material Figure S4). To quantify activity induced perturbations of the droplet interface, the rate of change of contact angle with time $\langle\frac{\Delta\theta}{\Delta t}\rangle$ was determined (Figure 4g) using the protocol described earlier.

At low fuel concentrations (0.1-1 wt%, Figures 4a-b), the temporal decay of the contact angle follows a largely monotonic trend. The droplet maintains a higher initial and intermediate contact angle due to reduced substrate-liquid affinity and enhanced curvature imposed by hydrophobic wettability. Although the overall trajectory resembles passive hydrophobic evaporation, small fluctuations are already evident, indicating that weak catalytic $O_2$ generation perturbs the interface. Although no visually resolved $O_2$ bubbles appear during the drying process, we speculate that some $O_2$ bubbles are present and subtly reduce the pinning duration through localised bubble-induced Marangoni stresses. At intermediate fuel concentrations (Figure 4c-d), evaporation departs markedly from smooth, monotonic behaviour. Following the initial pinned CCR stage, a distinct CCA regime is not established. Instead, the droplet transitions directly into a mixed mode. This behaviour is an indication of frequent bubble nucleation near the contact line. On the hydrophobic surface, particles and dissolved $O_2$ remain confined, promoting local supersaturation and delayed gas escape [45]. The resulting bubble growth induces localised interfacial deformation and spatially non-uniform surface tension gradients. These gradients generate bubble-induced Marangoni stresses that become comparable to the evaporation-driven capillary flux. Consequently, the contact line does not relax smoothly into a CCA regime but instead experiences repeated pinning-depinning-repinning events. The absence of a stable CCA phase therefore reflects the strong coupling between hydrophobic curvature, $O_2$ accumulation, and interfacial stress imbalance, which fragments the classical evaporation pathway into a dynamically mixed mode. At higher fuel concentrations (Figures 4e-f), evaporation becomes strongly oscillatory and temporally discontinuous. The contact angle graph displays large-amplitude, abrupt drops, indicating repeated cycles of bubble nucleation, growth, coalescence, and detachment at or near the liquid-air interface. Unlike the hydrophilic case (Figures 3d-f), where oscillations are comparatively small, the hydrophobic case exhibits larger and more frequent fluctuations, as directly observed in both the contact angle trajectories and side-view videos. These fluctuations are associated with intermittent bubble dynamics, which induce transient interfacial deformation and local stress imbalances. As a result, the contact line does not recede smoothly but undergoes successive perturbation events, each corresponding to localised relaxation following bubble growth or detachment. Thus, the evaporation trajectory consists of discrete relaxation episodes rather than a continuous decrease in curvature. In this regime, substrate wettability influences the frequency and persistence of bubble-induced perturbations.

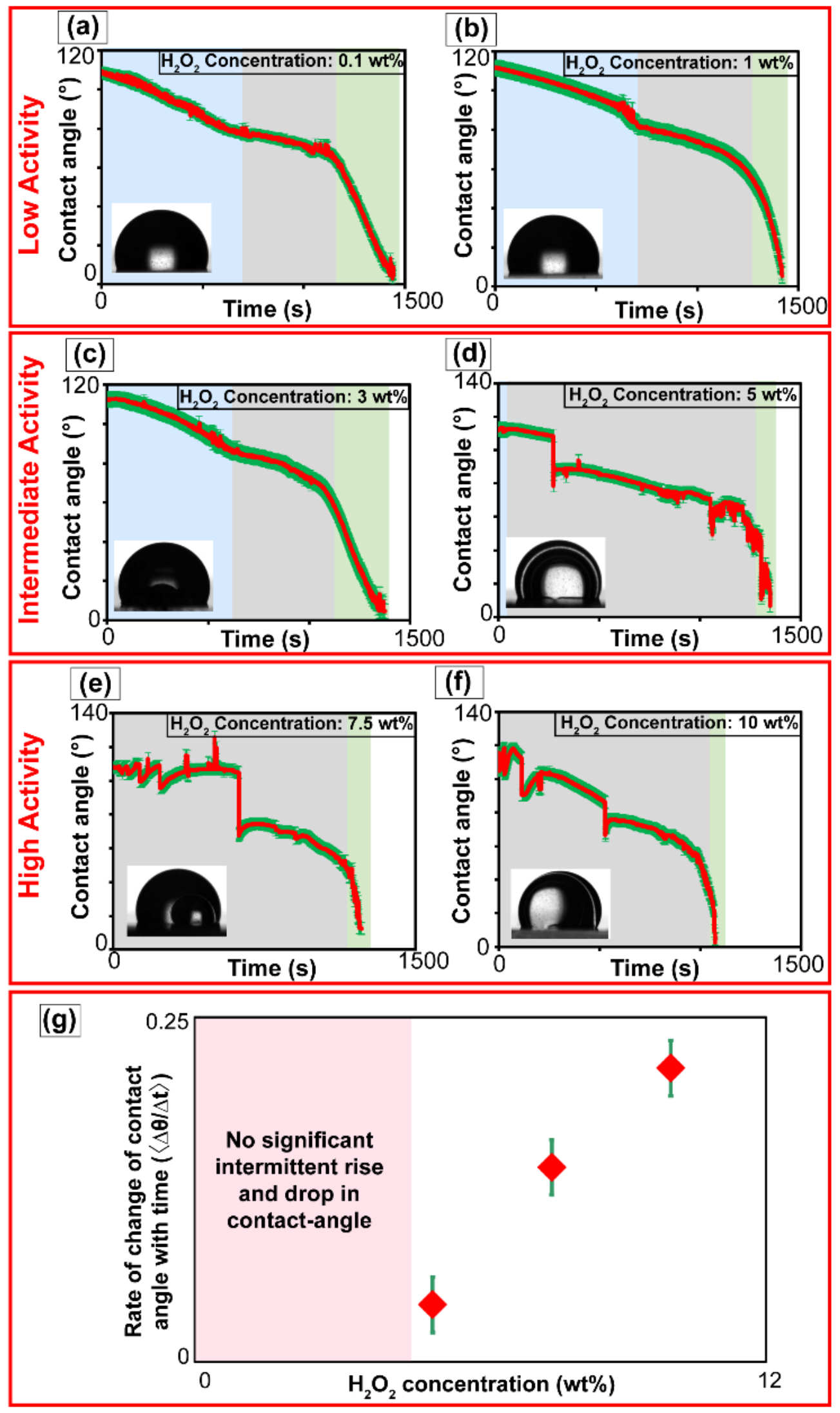

**Figure 4** Drying kinetics of catalytically active Janus particle droplets on a hydrophobic substrate. with increasing $H_2O_2$ concentration: (a) 0.1 wt%, (b) 1 wt%, (c) 3 wt%, (d) 5 wt%, (e) 7.5 wt%, and (f) 10 wt%. (a)-(f): Evolution of contact angle with time. The error bars calculated from five individual runs are highlighted in green. (g) The rate of change of contact angle with time $\langle\frac{\Delta\theta}{\Delta t}\rangle$.

$\langle\frac{\Delta\theta}{\Delta t}\rangle$ plotted against $H_2O_2$ concentration (Figure 4g) shows a steeper increase compared to hydrophilic substrates. indicating larger fluctuations due to higher $O_2$ bubble activity on the hydrophobic substrate. This provides a quantitative descriptor of the enhanced interfacial instability induced by the combined effects of catalytic $O_2$ generation and hydrophobicity.

***Influence of bubble lifetime on deposition morphologies of catalytically active Janus droplets:***

Figure 5 illustrates how the interplay between catalytic $O_2$ generation and vapour escape pathways governs particle organisation on both hydrophilic and hydrophobic substrates. Under open drying conditions on hydrophilic substrates (Figures 5a1,b1,c1), low fuel concentrations (0-1 wt%) produce deposits with a relatively thin but well-defined peripheral rings, reflecting sustained outward capillary transport and weak Marangoni stresses. The annulus shows mild corrugation due to local particle interactions and discrete pinning sites [10], while the high-magnification insets reveal limited clustering, indicating that bubble-mediated perturbations remain transient and do not significantly disrupt radial flux. In contrast, under closed drying (Figures 5a2,b2,c2), deposits become more spatially uniform at the same concentrations. The peripheral ring is less sharply defined and the interior particle density increases, consistent with restricted vapour diffusion, extended droplet lifetime, and reduced net outward capillary flow [19]. In contrast to the hydrophilic substrates discussed above, hydrophobic surfaces (Figures 5a3,b3,c3) produce thicker, broader, and more sharply defined peripheral rings under open conditions, consistent with stronger and prolonged contact line pinning that sustains outward advection. Insets show dense particle packing along the boundary with some interior clustering, confirming dominant evaporation-driven transport. Under closed conditions (Figures 5a4,b4,c4), however, the annular band becomes less distinct even on the hydrophobic substrates, reflecting reduced radial flux and inward recirculation. Overall, the deposit morphology reflects the weak influence of catalytic bubbles, with substrate-dependent pinning governing peripheral accumulation even as limited bubble-mediated recirculation only modestly redistributes particles under prolonged (closed) drying conditions.

At intermediate fuel concentrations (3-5 wt%) under open drying (Figures 5d1,e1), hydrophilic substrates retain a discernible peripheral ring, but it develops corrugations and local discontinuities due to repeated bubble nucleation-burst cycles. Bubble-induced Marangoni stresses intermittently weaken the capillary-driven outward flux [19,22]. Under closed conditions (Figures 5d2,e2), restricted vapour diffusion and longer bubble lifetimes enhance interfacial deformation and recirculation, producing broader, more diffuse deposits with reduced annular sharpness and greater interior particle density. On hydrophobic substrates (Figures 5d3,e3), the peripheral ring persists under open conditions but exhibits irregular thickening and corrugation from intermittent bubble-mediated pinning-depinning events. In closed systems (Figures 5d4,e4), prolonged bubble residence further destabilises the annular boundary, yielding a non-uniform and partially redistributed deposit due to sustained Marangoni recirculation and delayed evaporation. Overall, at intermediate fuel concentrations, prolonged bubble lifetime and enhanced Marangoni stresses increasingly compete with capillary transport, progressively destabilising the peripheral ring, which is amplified under closed drying conditions and moderated by stronger contact line pinning on hydrophobic substrates.

At high fuel concentrations (7.5-10 wt%), catalytic activity dominates the drying pathway and the contrast between open and closed environments becomes most pronounced. On hydrophilic substrates under open conditions (Figures 5f1,g1), rapid $O_2$ escape limits bubble lifetime, suppressing sustained Marangoni recirculation and smoothing the deposit into a more spatially uniform morphology with a weakened or disappearing peripheral ring. In closed systems (Figures 5f2,g2), restricted gas escape prolongs bubble lifetime and promotes repeated growth-coalescence cycles, generating persistent surface tension gradients and cyclic pinning-depinning events that produce multiple concentric bands instead of a single annulus. On hydrophobic substrates under open drying (Figures 5f3,g3), the coherent ring destabilises into discrete, radially oriented protrusions, forming a star-like morphology driven by frequent, short-lived Marangoni bursts that redirect particle transport along radial pathways. Under closed conditions (Figures 5f4,g4), prolonged bubble residence enhances sustained interfacial deformation, leading to fragmented annular segments and increased inward redistribution rather than smooth concentric rings. Overall, at high fuel concentrations, bubble lifetime emerges as the primary control parameter, with rapid gas escape in open systems favouring transient perturbations and uniform particle redistribution, while restricted escape in closed environments sustains interfacial deformation and drives complex, multi-banded or fragmented deposition patterns.

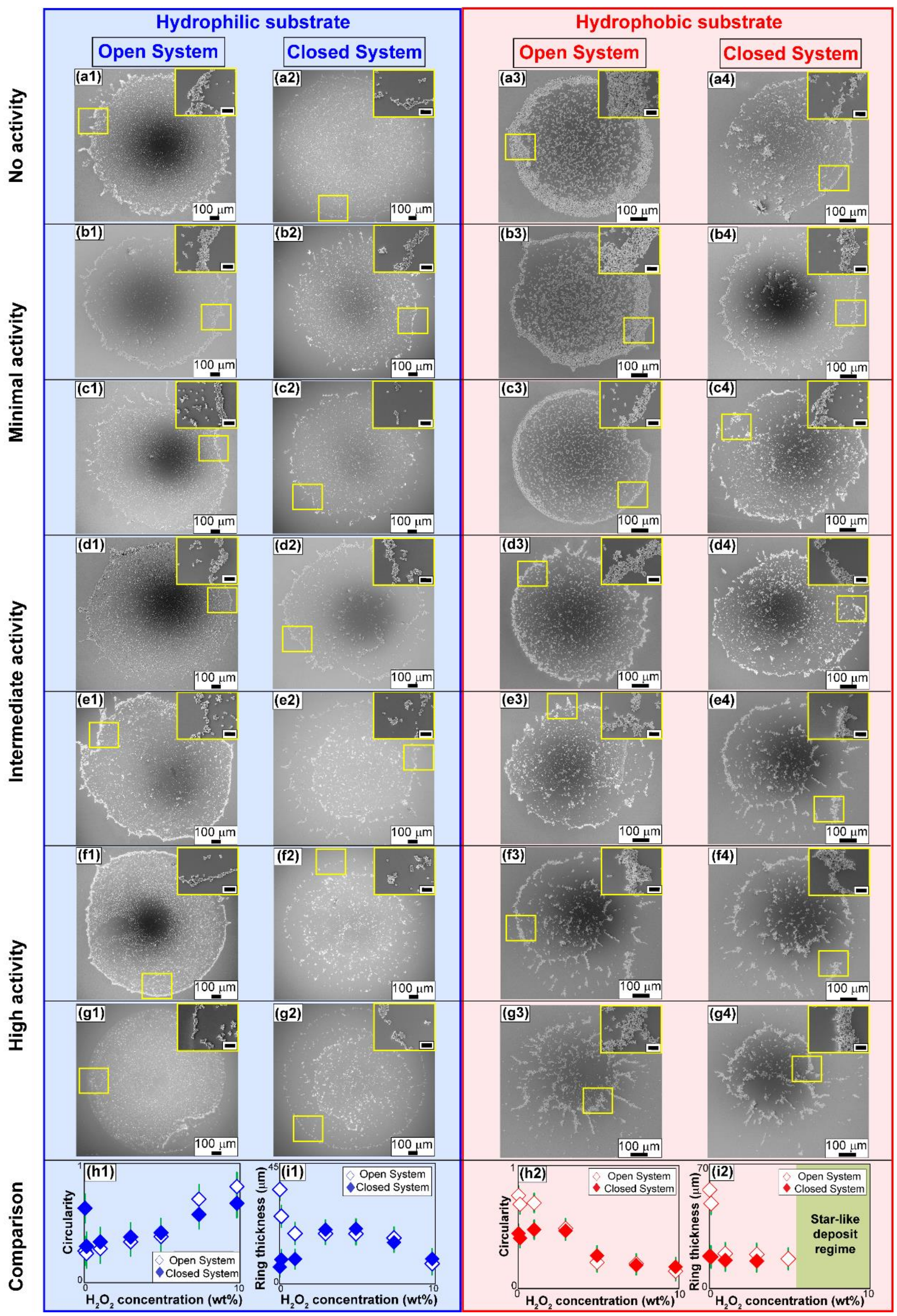

**Figure 5** SEM images of final dried droplet deposits on hydrophilic and hydrophobic surface with increasing $H_2O_2$ concentration: (a) 0 wt%, (b) 0.1 wt%, (c) 1 wt%, (d) 3 wt%, (e) 5 wt%, (f) 7.5 wt%, and (g) 10 wt%. (a1)-(g1), (a3)-(g3) respectively correspond to open drying conditions on hydrophilic and hydrophobic substrates. (a2)-(g2), (a4)-(g4) respectively correspond to closed drying conditions on hydrophilic and hydrophobic substrates. Insets show higher magnification views of selected regions to highlight microstructural details of the dried deposits. The scale bar in the inset is 20 μm. Variation in (h) circularity and (i) ring thickness of the deposit morphologies with increasing $H_2O_2$ concentration.

The morphological trends are quantitatively reflected in the circularity and ring-thickness analyses (Figures 5h-i). On hydrophilic substrates, circularity increases with $H_2O_2$ concentration in both environments (Figure 5h1), particularly under open conditions at high fuel levels, consistent with the observed suppression of a sharply defined annulus and emergence of more spatially uniform deposits. In closed systems, elevated circularity reflects sustained bubble-driven recirculation that redistributes particles while maintaining overall footprint symmetry. In contrast, on hydrophobic substrates (Figure 5h2), circularity decreases progressively with increasing fuel concentration in both open and closed environments, consistent with the development of radial protrusions and fragmented boundaries. This loss of symmetry is amplified under closed conditions due to prolonged bubble residence and stronger interfacial disturbances. Ring-thickness data further distinguish deposit uniformity across the droplet footprint. On hydrophilic substrates under open drying, ring thickness decreases monotonically with increasing fuel concentration (Figure 5i1), reflecting progressive weakening of sustained capillary-driven edge accumulation. In closed systems, a non-monotonic trend emerges, with intermediate concentrations showing temporary thickening near the centre of the deposit. This is attributed to persistent recirculation and pinning-depinning events. Eventually, multi-ring structures dominate at high activity. On hydrophobic substrates, even minimal fuel addition sharply reduces thickness in both environments (Figure 5i2), indicating rapid disruption of edge-directed advection. At higher concentrations, thickness remains relatively similar as catalytic activity suppresses stable annular build-up. Overall, catalytic activity shifts the system from capillarity-dominated toward bubble lifetime-controlled redistribution, with substrate wettability and vapour confinement determining whether deposit uniformity is preserved or progressively destabilised.

***Deposition heterogeneity revealed by AFM: role of wettability, activity, and evaporation environment***

Figure 6 provides nanoscale topographical insight into the spatial heterogeneity of PS-Pt active Janus particle deposits by resolving local height distributions and clustering patterns using AFM. Measurements were performed at both the edge and centre of dried droplets on hydrophilic and hydrophobic substrates, under open and closed evaporation environments, at representative low (1 wt%) and high (7.5 wt%) $H_2O_2$ concentrations. This mapping enables direct correlation between macroscopic deposit footprint morphology and particle-scale organisation.

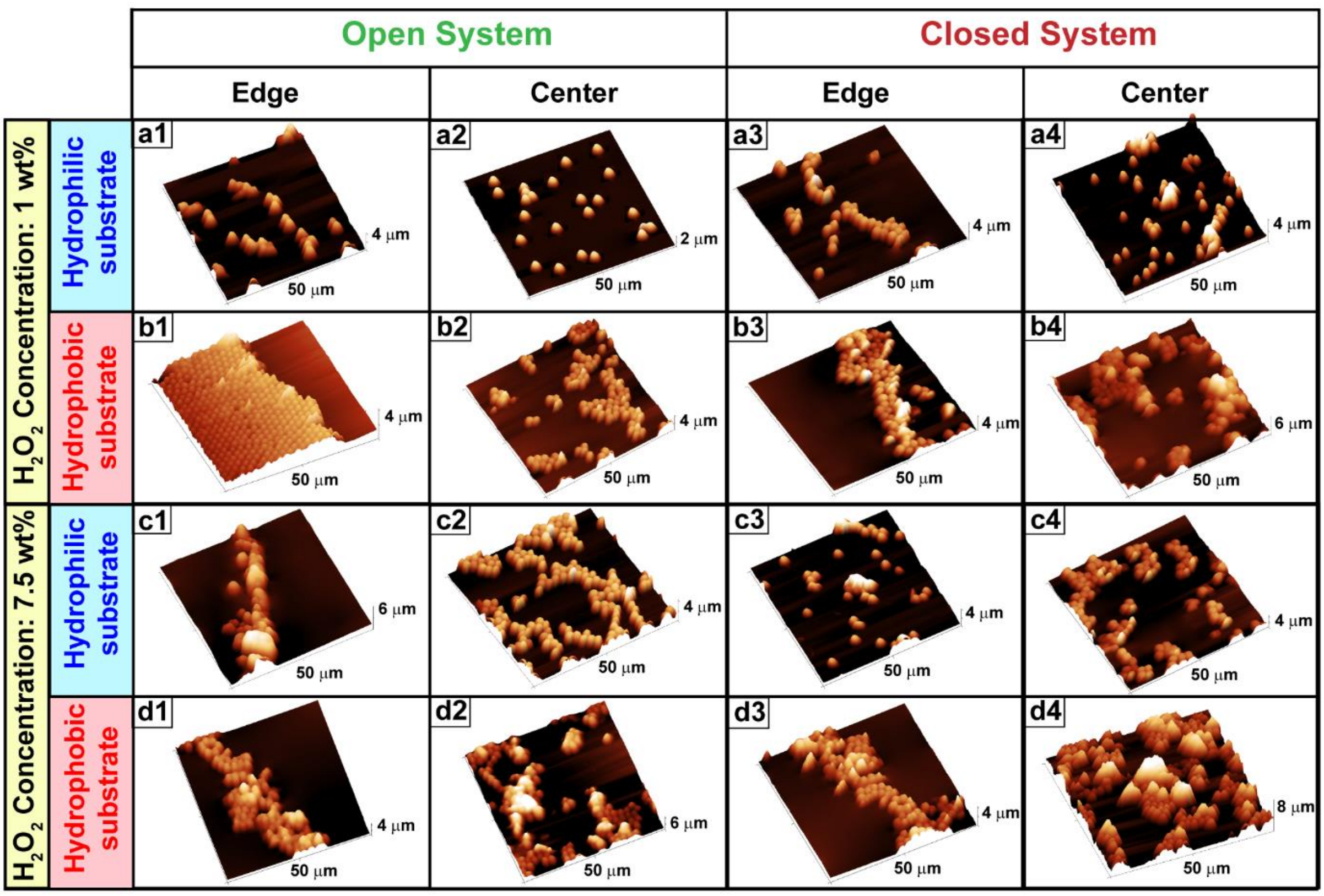

**Figure 6** Representative AFM images of dried droplet deposits on hydrophilic and hydrophobic substrates at different $H_2O_2$ concentrations: (a1-b4) 1 wt% and (c1-d4) 7.5 wt%. Rows represent substrate wettability: hydrophilic (a1-a4, c1-c4) and hydrophobic (b1-b4, d1-d4). Columns represent drying conditions: open (a1, a2, b1, b2, c1, c2, d1, d2) and closed (a3, a4, b3, b4, c3, c4, d3, d4). For each condition, AFM scans were taken at both the edge and centre of the droplet to capture spatial variations in deposition morphologies.

At low catalytic activity (1 wt% $H_2O_2$) on hydrophilic substrates, the AFM images under open conditions (Figures 6a1,a2) show relatively isolated particle islands at both the edge and centre, with limited lateral aggregation. The edge region displays slightly elongated clusters, whereas the centre retains sparse and evenly distributed particles. Under closed conditions (Figures 6a3,a4), both regions exhibit increased lateral clustering and broader height distributions, indicating that restricted vapour escape and prolonged droplet lifetime enhance particle-particle interactions. The difference between edge and centre becomes less pronounced in the latter case, reflecting partial redistribution driven by slower evaporation and mild Marangoni recirculation. On hydrophobic substrates at 1 wt% (Figures 6b1-b4), the heterogeneity is inherently stronger. Under open conditions, the edge region shows denser packing compared to the centre, reflecting prolonged contact line pinning and enhanced local accumulation. Under closed conditions, both edge and centre develop interconnected aggregates, with increased roughness contrast and reduced spatial separation between clusters. This indicates that unfavourable wetting amplifies boundary retention, while confinement promotes aggregation across the footprint rather than uniform spreading, because weaker solid-liquid affinity and higher interfacial curvature increase particle residence time near the interface and limit lateral stress relaxation.

At high catalytic activity (7.5 wt% $H_2O_2$), the AFM images reveal a marked transition from isolated deposits to continuous networked structures, indicating dominant bubble-mediated redistribution. On hydrophilic substrates under an open condition (Figures 6c1,c2), the edge region shows cluster coalescence, rather than isolated deposits as observed at low activity. The centre develops more interconnected cluster networks than observed at 1 wt% $H_2O_2$. The morphology on hydrophilic substrates under open condition reflects enhanced particle-particle interactions and intermittent redistribution driven by catalytic disturbances, but without clear directional alignment. Under closed conditions (Figures 6c3,c4), aggregation becomes more compact, particularly at the centre, where cluster coalescence reduces inter-cluster spacing. This arises from restricted vapour diffusion and prolonged bubble residence, which together extend the droplet lifetime and sustain bubble-induced Marangoni recirculation. The longer residence time increases the probability of particle-particle encounters and promotes repeated inward redistribution cycles, enabling neighbouring clusters to merge into larger aggregates rather than remain laterally dispersed [46,47]. In contrast, the edge region is characterised by discontinuous localised particle accumulation. On hydrophobic substrates at 7.5 wt%, heterogeneity reaches its maximum. Under open conditions (Figures 6d1,d2), the edge shows a uniform ring at some regions, while the centre exhibits clustered deposits reflecting rapid but spatially localised Marangoni bursts. Under closed conditions (Figures 6d3,d4), both edge and centre display densely packed, multi-level aggregates. The centre, in particular, develops structures indicative of repeated inward particle transport and coalescence. This morphology arises from the extended droplet lifetime and sustained internal flow loops in closed environments, which increase particle collision frequency and favour volumetric stacking rather than purely lateral spreading.

Overall, Figure 6 demonstrates that deposition heterogeneity is governed by the coupled action of wettability, catalytic activity and evaporation environment. Hydrophilic substrates at low activity favour isolated deposits, while hydrophobic substrates inherently amplify boundary clustering. Increasing catalytic activity transforms discrete islands into networked structures through bubble-induced Marangoni flows, while confinement further enhances vertical aggregation by extending bubble lifetime. The AFM analysis thus reveals that the transition from capillary-dominated to catalytically perturbed drying is manifested not only in global footprint morphology but also in cluster connectivity and edge-centre roughness contrast at the nanoscale.

***Mechanistic understanding and quantitative estimations of bubble lifetime and count***

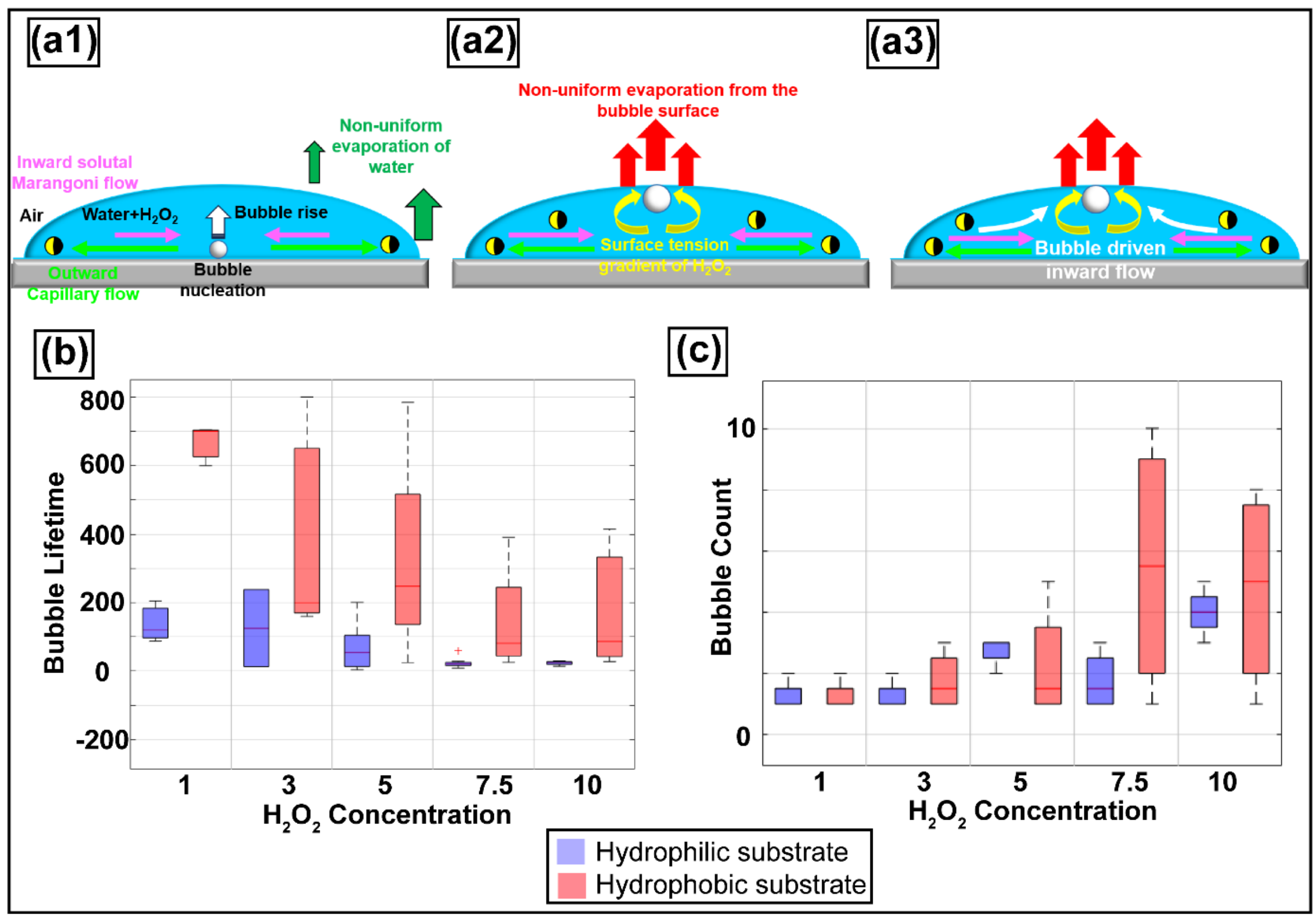

**Figure 7** (a1-a3) Schematic illustration of bubble-induced solutal Marangoni flow in catalytically active PS-Pt Janus droplets. (a1) Initial stage showing dominant outward capillary flow and onset of bubble nucleation, with the bubble rising to the top of the droplet. (a2) Bubble growth and localisation near the droplet apex induce non-uniform evaporation and $H_2O_2$ enrichment. (a3) Resulting surface tension gradients drive inward Marangoni flow (white arrows), which opposes capillary transport and redistributes particles toward the droplet centre. (b) Bubble lifetime as a function of $H_2O_2$ concentration on hydrophilic and hydrophobic substrates, showing a significantly stronger decrease with increasing fuel concentration for the hydrophobic substrate. (c) Corresponding variation in bubble count, indicating increased nucleation frequency with fuel concentration.

Figure 7a illustrates the proposed mechanism of bubble-induced inward transport in catalytically active droplets based on our observations and available literature [19,48,49,50]. In the initial stage (a1), evaporation-driven outward capillary flow dominates (shown by the light green arrows), leading to particle accumulation near the contact line, where catalytic decomposition of $H_2O_2$ progressively consumes fuel. As a result, the edge region becomes relatively depleted in $H_2O_2$. Pure $H_2O_2$ has a higher surface tension than water, due to which the edge has lower surface tension than the centre. This surface tension gradient drives an inward solutal Marangoni flow (shown by the pink arrows) [48]. With increasing activity, $O_2$ bubble nucleation occurs and bubbles migrate towards the droplet apex due to curvature and buoyancy (a2). Non-uniform evaporation along the bubble interface, with enhanced water loss near the apex, leads to local enrichment of $H_2O_2$ at the top of the bubble relative to its base, as shown in (a2). This generates a surface tension gradient along the bubble interface, with higher surface tension at the apex. A flow directed from the bubble base toward the apex is thereby established, which induces an inward recirculation within the droplet (shown by the yellow arrows in (a2) and (a3)). This flow, introduced as a bubble-induced inward Marangoni flow (a3), reinforces the inward solutal Marangoni flow (a1). Consequently, particles are advected towards the bubble and the droplet centre, while the outward capillary flux is weakened.

The influence of bubble dynamics on droplet behaviour is quantified in Figures 7b and 7c. Figure 7b presents the variation in bubble lifetime as a function of $H_2O_2$ concentration for hydrophilic and hydrophobic substrates. Bubble lifetime is consistently higher on hydrophobic substrates, indicating prolonged bubble residence due to reduced wettability and slower $O_2$ escape. In contrast, hydrophilic substrates exhibit shorter lifetimes due to enhanced spreading and faster relaxation of interfacial stresses. Figure 7c shows the corresponding bubble count, which increases with fuel concentration for both substrates, reflecting enhanced catalytic activity. The higher bubble count and longer lifetime on hydrophobic substrates together indicate stronger and more persistent interfacial perturbations, consistent with the observed deviations from capillary-dominated behaviour. However, resolving these quantities experimentally is challenging due to rapid contact line fluctuations and bubble coalescence.

**Conclusions**

This study identifies $O_2$ bubble lifetime as a control parameter governing evaporation dynamics, interfacial stability, and final deposition morphology of catalytically active Janus particle droplets on flat surfaces. Through systematic variations of hydrogen peroxide concentration, evaporation environment and surface wettability, we show that active droplet drying is regulated by a set of coupled mechanisms. These mechanisms are evaporation-driven capillary transport, Marangoni stress relaxation, and catalytic $O_2$ generation.

At low activity, evaporation-controlled capillary flows dominate on both hydrophilic and hydrophobic substrates, yielding smooth contact angle decay and relatively ordered peripheral deposits. With increasing activity, catalytic forces progressively perturb the interfacial stress balance, introducing intermittent oscillations and mixed transport regimes characterised by competing inward and outward flows. Beyond a critical activity threshold, bubble-mediated disturbances become the primary driver of interfacial dynamics, producing rapid curvature fluctuations, shortened pinning intervals, and increasingly heterogeneous deposition patterns. Hydrophilic substrates mitigate these effects through enhanced wetting and faster stress relaxation, whereas hydrophobic substrates prolong interfacial bubble lifetime and $O_2$ retention, thereby lowering the instability threshold and amplifying particle disorder.

Time-resolved contact angle analysis with structural characterisation also establishes the key role of bubble growth and burst cycles as the decisive parameter that governs the transition from capillary-dominated to catalytically perturbed drying in active Janus droplets. The rate of contact angle change and deposit circularity emerge as robust quantitative descriptors of interfacial destabilisation, linking dynamic evaporation pathways to final deposition outcomes.

A comparison table summarising published results on passive and active interfaces has been included in Supplementary Material (**Table S1**). Our results recast evaporating colloidal Janus droplets as chemically responsive interfaces rather than passive transport systems. The design framework developed here provides practical guidelines for tuning pattern formation in reactive coatings, printing, and microreactor platforms, while contributing to a deeper understanding of nonequilibrium interfacial phenomena in active colloidal matter. Since a regime characterised by strongly-coupled reaction-driven bubble generation, evaporation, and interfacial transport has not yet been treated theoretically, the development of a predictive model remains an open challenge and represents an important direction for future work. We believe that our experimental findings provide an essential foundation for such theoretical developments.

**Acknowledgements:** We would like to thank Mr. Yatheendran K. M. for assistance with SEM and AFM imaging.

**Research ethics:** Not applicable.

**Informed consent:** Not applicable.

**Author contributions:** Meneka Banik: Conceptualisation, Data curation, Methodology, Validation, Formal analysis, Investigation, Writing - Original Draft, Visualisation.

Ranjini Bandyopadhyay: Methodology, Validation, Resources, Data curation, Writing - Review & Editing, Supervision, Project administration, Funding acquisition.

**Use of Large Language Models, AI and Machine Learning Tools:** None declared.

**Conflict of interest:** The authors state no conflict of interest.

**Research funding:** We would like to thank Raman Research Institute, Bangalore for funding the research.

**Data availability:** The data that support the findings of this study are available on request.

**References**

1. R. D. Deegan, O. Bakajin, T. F. Dupont, G. Huber, S. R. Nagel, and T. A. Witten, "Capillary flow as the cause of ring stains from dried liquid drops," *Nature*, vol. 389, no. 6653, pp. 827–829, 1997.
2. R. D. Deegan, "Pattern formation in drying drops," *Phys. Rev. E*, vol. 61, no. 1, pp. 475–485, 2000.
3. H. Hu and R. G. Larson, "Evaporation of a sessile droplet on a substrate," *J. Phys. Chem. B*, vol. 106, no. 6, pp. 1334–1344, 2002.
4. R. G. Larson, "Transport and deposition patterns in drying sessile droplets," *AIChE J.*, vol. 60, no. 5, pp. 1538–1571, 2014.
5. H. Hu and R. G. Larson, "Marangoni effect reverses coffee-ring depositions," *J. Phys. Chem. B*, vol. 110, no. 14, pp. 7090–7097, 2006.
6. W. D. Ristenpart, P. G. Kim, C. Domingues, J. Wan, and H. A. Stone, "Influence of substrate conductivity on circulation reversal in evaporating drops," *Phys. Rev. Lett.*, vol. 99, pp. 234502, 2007.
7. K. Sefiane, "Patterns from drying drops," *Adv. Colloid Interface Sci.*, vol. 206, pp. 372–381, 2014.
8. B. Derby, "Inkjet printing of functional and structural materials: Fluid property requirements, feature stability, and resolution," *Annu. Rev. Mater. Res.*, vol. 40, pp. 395–414, 2010.
9. J. Park and J. Moon, "Control of colloidal particle deposit patterns within picoliter droplets ejected by ink-jet printing," *Langmuir*, vol. 22, no. 8, pp. 3506–3513, 2006.
10. P. J. Yunker, T. Still, M. A. Lohr, and A. G. Yodh, "Suppression of the coffee-ring effect by shape-dependent capillary interactions," *Nature*, vol. 476, no. 7360, pp. 308–311, 2011.
11. S. Michelin and E. Lauga, "Phoretic self-propulsion at finite Péclet numbers," *J. Fluid Mech.*, vol. 747, pp. 572–604, 2014.

12. J. L. Moran and J. D. Posner, "Phoretic self-propulsion," *Annu. Rev. Fluid Mech.*, vol. 49, pp. 511–540, 2017.
13. A. Walther and A. H. E. Müller, "Janus particles: Synthesis, self-assembly, physical properties, and applications," *Chem. Rev.*, vol. 113, no. 7, pp. 5194–5261, 2013.
14. B. P. Binks and P. D. I. Fletcher, "Particles adsorbed at the oil–water interface: A theoretical comparison between spheres of uniform wettability and Janus particles," *Langmuir*, vol. 17, no. 16, pp. 4708–4710, 2001.
15. J. Zhang, B. A. Grzybowski, and S. Granick, "Janus particle synthesis, assembly, and application," *Langmuir*, vol. 33, no. 28, pp. 6964–6977, 2017.
16. M. Banik, S. Sett, C. Bakli, A. K. Raychaudhuri, S. Chakraborty, and R. Mukherjee, "Substrate wettability guided oriented self-assembly of Janus particles," *Sci. Rep.*, vol. 11, pp. 1182, 2021.
17. Q. Xie, G. B. Davies, and J. Harting, "Direct assembly of magnetic Janus particles at a droplet interface," *ACS Nano*, vol. 11, no. 11, pp. 11232–11239, 2017.
18. S. Jiang and S. Granick, "Janus balance of amphiphilic colloidal particles," *J. Chem. Phys.*, vol. 127, pp. 161102, 2007.
19. K. Singh, P. Kumar, H. Raman, H. Sharma, and R. Mangal, "Tailoring the coffee ring effect by chemically active Janus colloids," *ACS Appl. Eng. Mater.*, vol. 3, pp. 275–285, 2025.
20. W. E. Uspal, M. N. Popescu, S. Dietrich, and M. Tasinkevych, "Self-propulsion of a catalytically active particle near a planar wall," *Soft Matter*, vol. 11, pp. 434–438, 2015.
21. S. Thutupalli, R. Seemann, and S. Herminghaus, "Swarming behavior of simple model squirmers," *New J. Phys.*, vol. 13, pp. 073021, 2011.
22. M. Banik and R. Bandyopadhyay, "Bubble-driven flow transitions in evaporating active droplets on structured surfaces," *arXiv* 2511.22423, 2025, https://doi.org/10.48550/arXiv.2511.22423.
23. R. Malinowski, G. Volpe, I. P. Parkin, and G. Volpe, "Dynamic control of particle deposition in evaporating droplets," *J. Phys. Chem. Lett.*, vol. 9, no. 3, pp. 659–664, 2018.
24. I. Buttinoni, J. Bialké, F. Kümmel, H. Löwen, C. Bechinger, and T. Speck, "Dynamical clustering and phase separation in suspensions of self-propelled colloidal particles," *Phys. Rev. Lett.*, vol. 110, pp. 238301, 2013.

25. T. Andac, P. Weigmann, S. K. P. Velu, E. Pince, G. Volpe, G. Volpe, and A. Callegari, "Active matter alters the growth dynamics of coffee rings," *Soft Matter*, vol. 15, pp. 1488–1496, 2019.
26. S. J. Ebbens and J. R. Howse, "In pursuit of propulsion at the nanoscale," *Soft Matter*, vol. 6, no. 4, pp. 726–738, 2010.
27. S. Das, A. Garg, A. I. Campbell, J. Howse, A. Sen, D. Velegol, R. Golestanian, and S. J. Ebbens, "Boundaries can steer active Janus spheres," *Nat. Commun.*, vol. 6, pp. 8999, 2015.
28. H. Hu and R. G. Larson, "Analysis of the microfluid flow in an evaporating sessile droplet," *Langmuir*, vol. 21, no. 9, pp. 3963–3971, 2005.
29. D. Mampallil and H. B. Eral, "A review on suppression and utilization of the coffee-ring effect," *Adv. Colloid Interface Sci.*, vol. 252, pp. 38–54, 2018.
30. H. Y. Erbil, "Evaporation of pure liquid sessile and spherical suspended drops: A review," *Adv. Colloid Interface Sci.*, vol. 170, pp. 67–86, 2012.
31. D. Bonn, J. Eggers, J. Indekeu, J. Meunier, and E. Rolley, "Wetting and spreading," *Rev. Mod. Phys.*, vol. 81, pp. 739–805, 2009.
32. P. G. de Gennes, "Wetting: statics and dynamics," *Rev. Mod. Phys.*, vol. 57, pp. 827–863, 1985.
33. R. G. Picknett and R. Bexon, "The evaporation of sessile or pendant drops in still air," *J. Colloid Interface Sci.*, vol. 61, pp. 336–350, 1977.
34. K. S. Birdi, D. T. Vu, and A. Winter, "The evaporation of a liquid drop on a solid surface," *J. Phys. Chem.*, vol. 93, pp. 3702–3703, 1989.
35. J. R. Howse, R. A. L. Jones, A. J. Ryan, T. Gough, R. Vafabakhsh, and R. Golestanian, "Self-motile colloidal particles: From directed propulsion to random walk," *Phys. Rev. Lett.*, vol. 99, pp. 048102, 2007.
36. J. G. Gibbs and Y.-P. Zhao, "Autonomously motile catalytic nanomotors by bubble propulsion," *Appl. Phys. Lett.*, vol. 94, pp. 163104, 2009.
37. H.-J. Butt, J. Liu, K. Koynov, B. Straub, C. Hinduja, I. Roismann, R. Berger, X. Li, D. Vollmer, W. Steffen, and M. Kappl, "Contact angle hysteresis," *Curr. Opin. Colloid Interface Sci.*, vol. 59, pp. 101574, 2022.
38. R. Adkins, I. Kolvin, Z. You, S. Witthaus, M. C. Marchetti, and Z. Dogic, "Dynamics of active liquid interfaces," *Science*, vol. 377, no. 6606, pp. 768–772, 2022.
39. E. B. Dussan, "On the spreading of liquids on solid surfaces: Static and dynamic contact lines," *Annu. Rev. Fluid Mech.*, vol. 11, pp. 371–400, 1979.

40. A. Mozaffari, N. Sharifi-Mood, J. Koplik, and C. Maldarelli, “Self-diffusiophoretic colloidal propulsion near a solid boundary,” *Phys. Fluids*, vol. 28, pp. 053107, 2016.
41. N. D. Denkov, O. D. Velev, P. A. Kralchevsky, I. B. Ivanov, H. Yoshimura, and K. Nagayama, “Mechanism of formation of two-dimensional crystals from latex particles on substrates,” *Langmuir*, vol. 8, no. 12, pp. 3183–3190, 1992.
42. A. Perro, S. Reculusa, S. Ravaine, E. Bourgeat-Lami, and E. Duguet, “Design and synthesis of Janus micro- and nanoparticles,” *J. Mater. Chem.*, vol. 15, pp. 3745–3760, 2005.
43. M. Banik and R. Shenhar, “Nanoparticle assembly by transient topography induced by applying soft lithography to block copolymer films,” *Soft Matter*, vol. 20, pp. 4035–4042, 2024.
44. B. M. Weon and J. H. Je, “Self-pinning by colloids confined at a contact line,” *Phys. Rev. Lett.*, vol. 110, pp. 028303, 2013.
45. A. Al-Sharafi, B. S. Yilbas, A. Z. Sahin, H. Ali, and H. Al-Qahtani, “Heat transfer characteristics and internal fluidity of a sessile droplet on hydrophilic and hydrophobic surfaces,” *Appl. Therm. Eng.*, vol. 108, pp. 628–640, 2016.
46. I. Theurkauff, C. Cottin-Bizonne, J. Palacci, C. Ybert, and L. Bocquet, “Dynamic clustering in active colloidal suspensions with chemical signaling,” *Phys. Rev. Lett.*, vol. 108, pp. 268303, 2012.
47. M. Jalaal, B. ten Hagen, H. Le The, C. Diddens, D. Lohse, and Á. Marin, “Interfacial aggregation of self-propelled Janus colloids in sessile droplets,” *Phys. Rev. Fluids*, vol. 7, pp. 110514, 2022.
48. H. Kim, K. Muller, O. Shardt, S. Afkhami, and H. A. Stone, “Solutal Marangoni flows of miscible liquids drive transport without surface contamination,” *Nat. Phys*., vol. 13, pp. 1105-1110, 2017.
49. M. Manjare, F. Yang, R. Qiao, and Y. Zhao, “Marangoni flow induced collective motion of catalytic micromotors,” *J. Phys. Chem. C*, vol. 119, pp. 28361−28367, 2015.
50. P. Sharan, A. Daddi-Moussa-Ider, J. Agudo-Canalejo, R. Golestanian, and J. Simmchen, “Pair Interaction between Two Catalytically Active Colloids,” *Small*, vol. 19, pp. 2300817, 2023.

# Supplementary Material

# Influence of Bubble Lifetime on the Drying of Catalytically Active Sessile Droplets

Meneka Banik and Ranjini Bandyopadhyay*

Soft Condensed Matter Group, Raman Research Institute, C. V. Raman Avenue, Sadashivanagar, Bangalore 560080, India.

Corresponding author: *ranjini@rri.res.in

## *TABLE S1 COMPARISON TABLE:*

The table below summarises prior studies on passive and active droplet systems, emphasising how interfacial driving forces evolve in these systems. It highlights the lack of a unified framework linking wettability, bubble dynamics, and deposition morphology, which is addressed in the present work.

**Table S1** Comparison of passive and active droplet systems highlighting the dominant interfacial driving mechanisms, roles of Marangoni stresses and bubbles, and resulting deposition outcomes. The table identifies the key scientific gap addressed in this work.

| **System** | **Ref.** | **Dominant interfacial driving mechanism** | **Role of Marangoni stresses** | **Role of bubbles** | **Key conceptual outcome** |
|---|---|---|---|---|---|
| Passive colloidal droplets on flat substrates | **[1,2]** | Evaporation-driven outward capillary flow | Weak; thermal or solutal inward stresses | Absent | Coffee-ring formation |
| Passive droplets on patterned or heterogeneous substrates | **[3]** | Outward capillary flow with modified pinning | May be locally altered by surface chemistry | Absent | Deposition morphology modified |
| Active Janus particles in bulk or confined liquid (non-evaporating) | **[4]** | Self-diffusiophoresis / bubble propulsion | Local, particle scale | Central to propulsion and mixing | Activity alters particle motion |
| Active droplets on flat hydrophilic substrates | **[5]** | Capillary flow with activity induced perturbations | Present but spatially unregulated | Bubbles nucleate and disrupt local deposition | Bubble activity introduces local disorder |
| Active droplets with motile microorganisms (bacteria) under evaporation | **[6]** | Capillary flow coupled with motility-induced particle transport | Not explicitly dominant | Absent | Activity becomes relevant when motility competes with evaporation-driven transport |
| Active droplets on micro- and nano-patterned and wettability-tuned substrates | **[7]** | Capillary flow coupled with activity-induced Marangoni stress | Droplet-scale and coherent on micro-patterns; spatially localised on nano-patterns; reversed on hydrophilic micro-patterns | Bubbles grow, migrate and collapse on micro-patterns; remain pinned on nano-patterns; suppressed on hydrophilic micro-patterns | Topography and wettability determine drying patterns and flow regimes |
| Active droplets on flat substrates with tunable wettability and evaporation environment | **This work** | Capillary flow coupled with activity-induced interfacial perturbations governed by bubble residence time | Transient and spatially evolving; arise from local compositional gradients near the interface and are modulated by bubble persistence | Bubble nucleation, growth, and residence time regulate transport pathways | Bubble residence time emerges as an effective parameter linking catalytic activity, wettability, and evaporation conditions |

## *FIGURE S1: FABRICATION OF JANUS PARTICLES*

The fabrication process involves the assembly of a close-packed PS colloidal monolayer on a glass substrate, followed by Pt deposition to achieve a hemispherical coating. Subsequent sonication releases the particles into suspension without compromising their Janus structure.

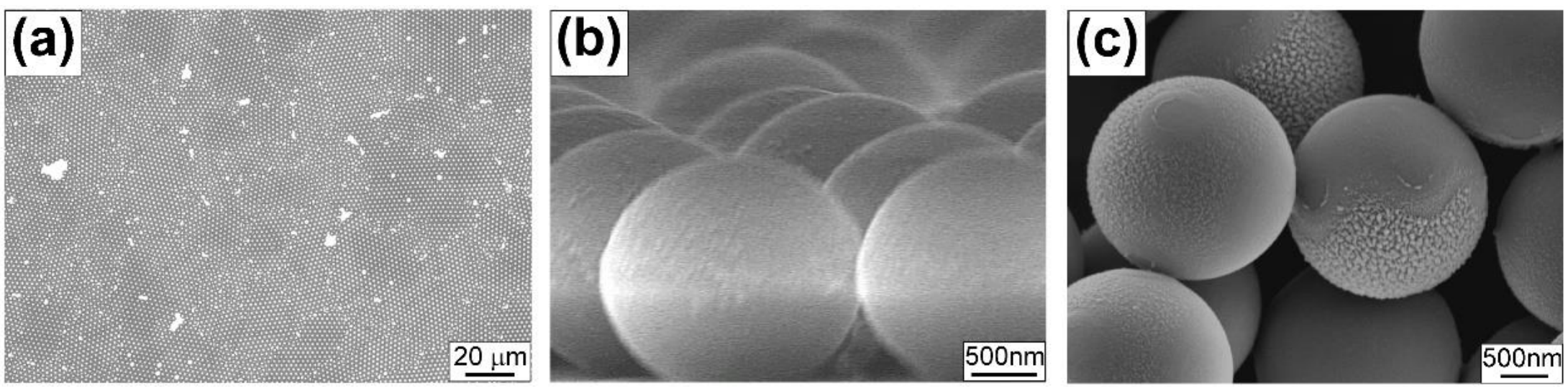

**Figure S8** Fabrication and morphology of PS-Pt Janus particles. (a) Optical microscope image showing a large-area close-packed monolayer of polystyrene (PS) colloids assembled on a glass substrate prior to metal deposition. (b) Side-view magnified SEM image illustrating partial platinum (Pt) coating on the exposed hemispheres of the PS particles via sputter deposition. (c) SEM image of the resulting PS-Pt Janus particles after sonication-assisted release from the substrate, confirming anisotropic surface morphology and successful Janus architecture.

## *FIGURE S2: ROUGHNESS OF THE FLAT SUBSTRATES.*

Atomic force microscopy (AFM) measurements were performed to quantify the surface roughness of the hydrophilic (SiW) and hydrophobic (PDMS) substrates prior to droplet deposition.

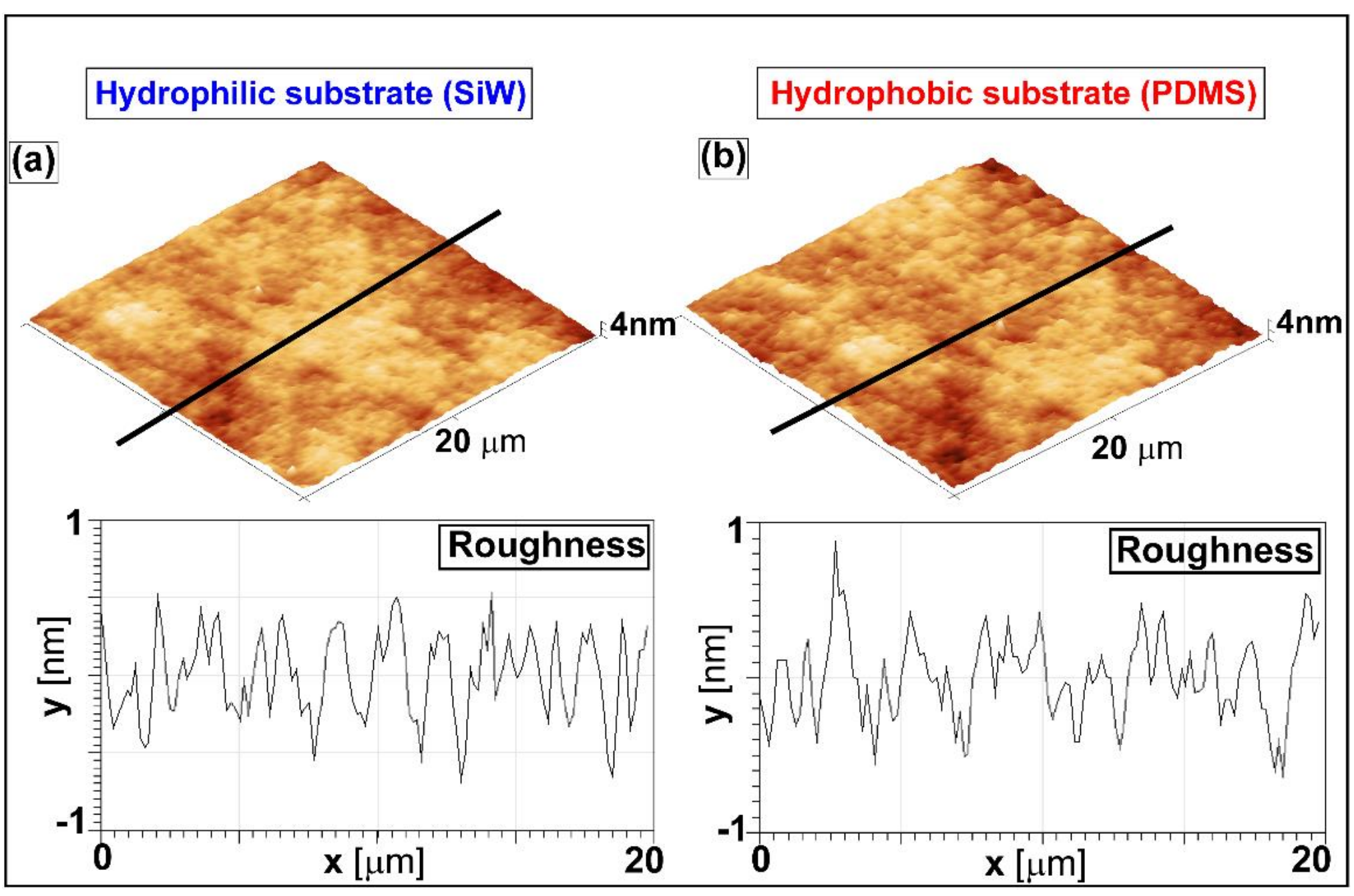

**Figure S9** AFM topographical characterisation of flat hydrophilic and hydrophobic substrates used in evaporation experiments. (a) Three-dimensional AFM image of the hydrophilic Si wafer (SiW) surface, along with the corresponding linear profiles extracted along the indicated directions. (b) AFM height image of the hydrophobic PDMS substrate with its corresponding linear profile. Both show nanoscale height fluctuations.

### *FIGURE S3: VARIATION IN BUBBLE NUCLEATION AND GROWTH WITH FUEL CONCENTRATION ON HYDROPHILIC SUBSTRATES*

With increasing $H_2O_2$ concentration, bubble nucleation becomes more frequent and bubbles persist longer near the droplet interface. Their preferential localisation and growth enhance interfacial deformation and generate stronger bubble-induced Marangoni flows, leading to increased disruption of capillary transport and progressively irregular deposition behaviour.

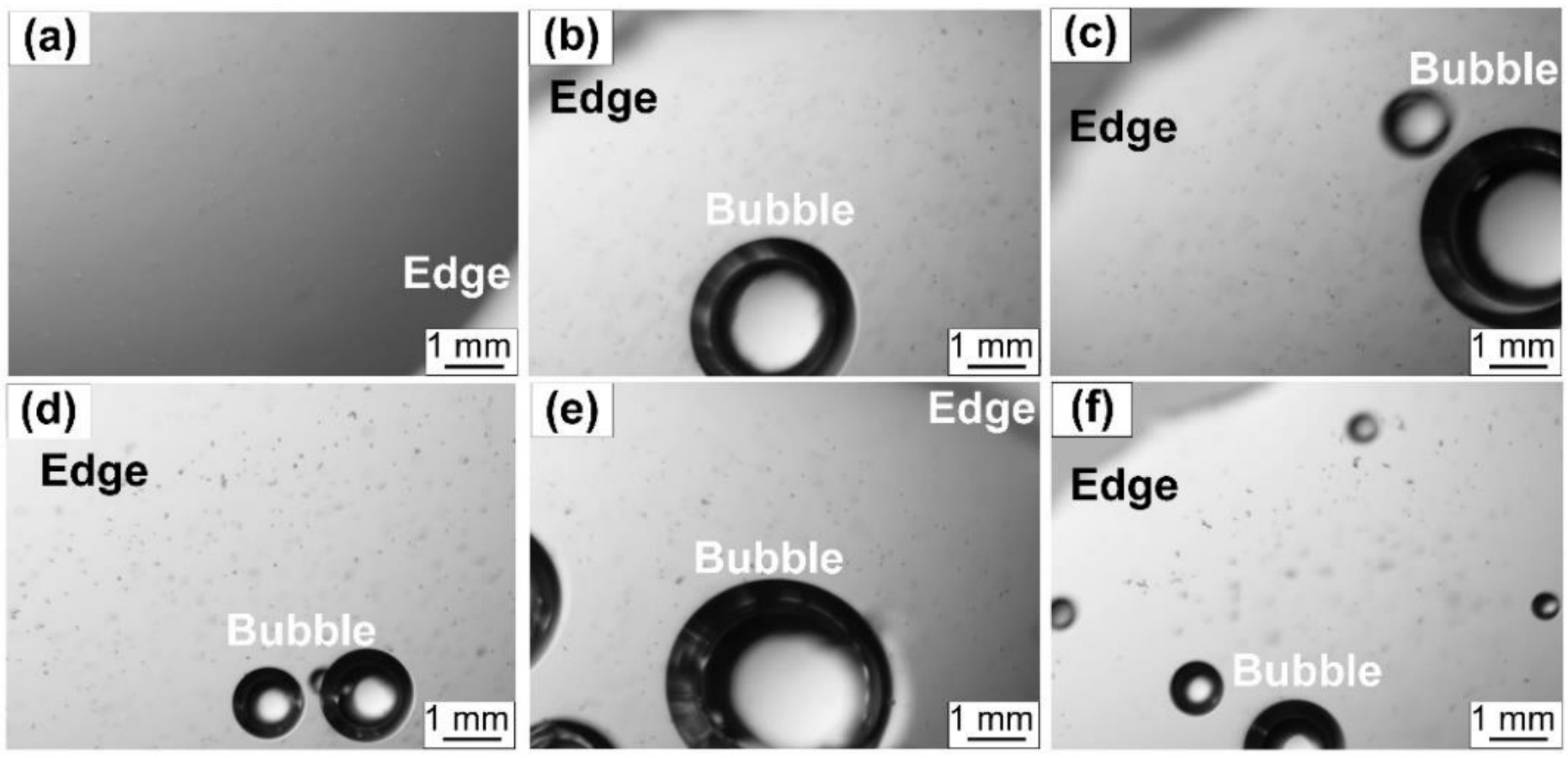

**Figure S10** Top-view optical microscope images captured at intermediate time, showing enhanced evolution of $O_2$ bubbles in catalytically active Janus particle droplets on a flat hydrophilic substrate with increasing $H_2O_2$ concentration: (a) 0.1 wt%, (b) 1 wt%, (c) 3 wt%, (d) 5 wt%, (e) 7.5 wt%, and (f) 10 wt%.

### *FIGURE S4: VARIATION IN BUBBLE NUCLEATION AND GROWTH WITH FUEL CONCENTRATION ON HYDROPHOBIC SUBSTRATES*

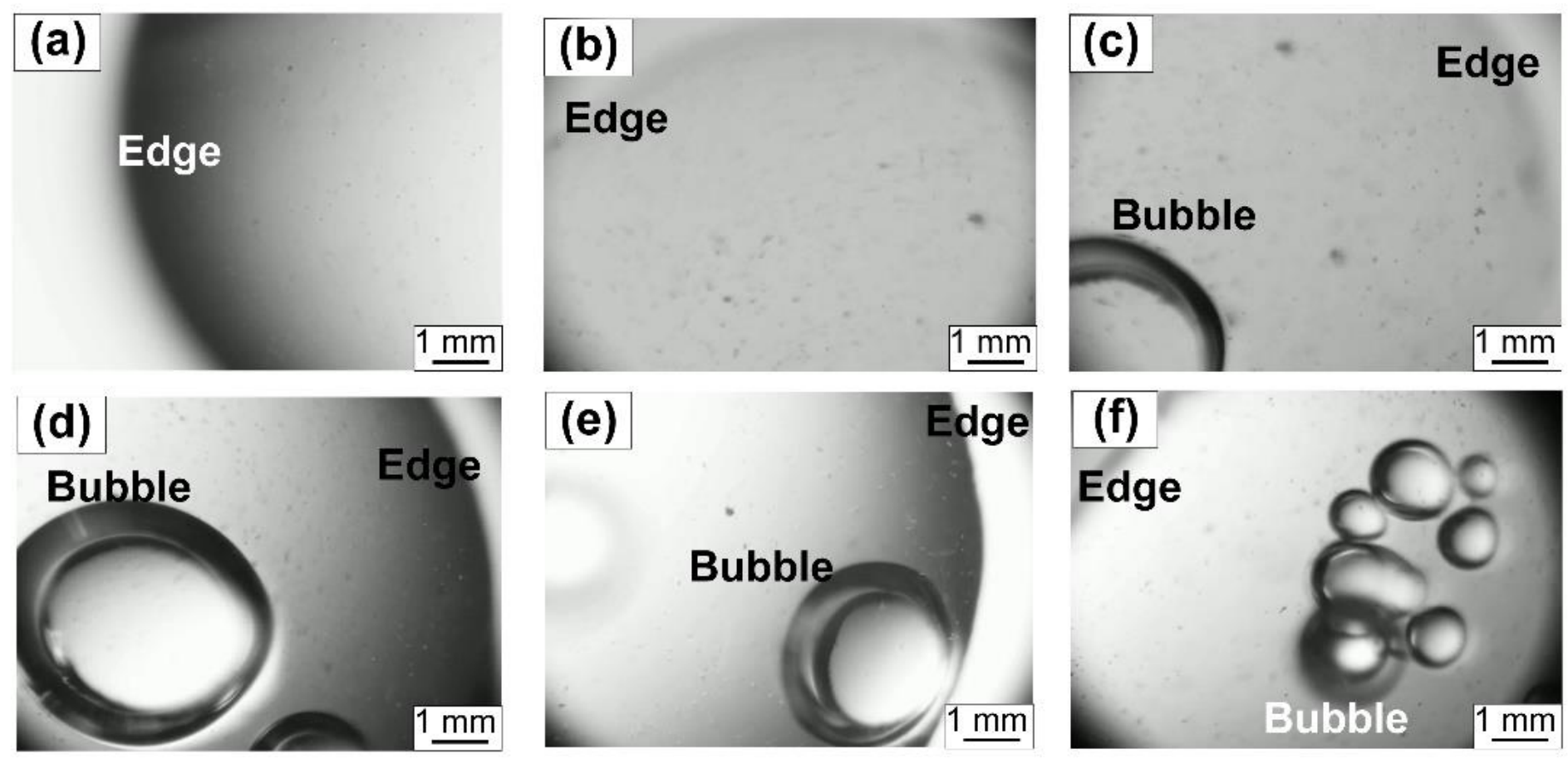

**Figure S11** Top-view optical microscope images captured at intermediate time, showing enhanced evolution of $O_2$ bubbles in catalytically active Janus particle droplets on a flat hydrophobic substrate with increasing $H_2O_2$ concentration: (a) 0.1 wt%, (b) 1 wt%, (c) 3 wt%, (d) 5 wt%, (e) 7.5 wt%, and (f) 10 wt%.

**References:**

1. R. D. Deegan, O. Bakajin, T. F. Dupont, G. Huber, S. R. Nagel, and T. A. Witten, "Capillary flow as the cause of ring stains from dried liquid drops," *Nature*, vol. 389, no. 6653, pp. 827–829, 1997.
2. R. D. Deegan, "Pattern formation in drying drops," *Phys. Rev. E*, vol. 61, no. 1, pp. 475–485, 2000.
3. D. Lohani, M. G. Basavaraj, D. K. Satapathy, and S. Sarkar, "Coupled effect of concentration, particle size and substrate morphology on the formation of coffee rings", Colloids Surf. A Physicochem. Eng. Asp, vol. 589, no. 20, 124387, 2020.
4. J. G. Gibbs and Y.-P. Zhao, "Autonomously motile catalytic nanomotors by bubble propulsion," *Appl. Phys. Lett.*, vol. 94, pp. 163104, 2009.
5. K. Singh, P. Kumar, H. Raman, H. Sharma, and R. Mangal, "Tailoring the coffee ring effect by chemically active Janus colloids," *ACS Appl. Eng. Mater.*, vol. 3, pp. 275–285, 2025.
6. T. Andac, P. Weigmann, S. K. P. Velu, E. Pince, G. Volpe, G. Volpe, and A. Callegari, "Active matter alters the growth dynamics of coffee rings," *Soft Matter*, vol. 15, pp. 1488–1496, 2019.
7. M. Banik and R. Bandyopadhyay, "Bubble-driven flow transitions in evaporating active droplets on structured surfaces," *arXiv* 2511.22423, 2025, https://doi.org/10.48550/arXiv.2511.22423.